# Mitigation of MHD induced fast-ion redistribution in MAST and implications for MAST-Upgrade design

D L Keeling[1], T R Barrett[1], M Cecconello[2], C D Challis[1], N Hawkes[1], O M Jones[1,3], I Klimek[2], K G McClements[1], A Meakins[1], J Milnes[1], M Turnyanskiy[4] and the MAST team[1]

[1] CCFE, Culham Science Centre, Abingdon, Oxon., OX14 3DB, UK
[2] Department of Physics and Astronomy, Uppsala University, SE-751 05 Uppsala, Sweden
[3] Department of Physics, Durham University, South Road, Durham DH1 3LE, UK
[4] ITER Physics Department, EFDA CSU Garching, Boltzmannstraße 2, D-85748 Garching, Germany

E-mail: David.Keeling@ccfe.ac.uk

## Abstract

The phenomenon of redistribution of neutral beam fast-ions due to MHD activity in plasma has been observed on many tokamaks and more recently has been a focus of research on MAST (Turnyanskiy M. *et al*, 2011 *Nucl. Fusion* **53** 053016). n=1 fishbone modes are observed to cause a large decrease in the neutron emission rate corresponding to a significant perturbation of the fast-ion population in the plasma. Theoretical work on fishbone modes states that the fast-ion distribution itself acts as the source of free energy driving the modes that cause the redistribution. Therefore a series of experiments have been carried out on MAST to investigate a range of plasma density levels at two neutral beam power levels to determine the region within this parameter space in which MHD activity and consequent fast-ion redistribution is suppressed. Analysis of these experiments shows complete suppression of MHD activity at high density with increasing activity and fast-ion redistribution at lower densities and higher NB power accompanied by strong evidence for localisation of the redistribution effect to a specific region in the plasma core. The results also indicate correlations between the form of the modelled fast-ion distribution function, the amplitude and growth rate of the fishbone modes, and the magnitude of the redistribution effect. The same analysis has been carried out on models of MAST-Upgrade baseline plasma scenarios to determine whether significant fast-ion redistribution is likely to occur in that device. A simple change to the neutral-beam injector geometry is proposed which is shown to have a significant mitigating effect in terms of the fishbone mode drive and is therefore expected to allow effective plasma heating and current drive over a wider range of plasma conditions in MAST-Upgrade.

## 1. INTRODUCTION

The Neutral Beam injection (NBI) system on MAST consists of two injection devices capable of injecting neutral deuterium with energies up to 75keV and total power exceeding 2MW per injector [1]. NBI has been used extensively in recent MAST experimental campaigns to access high performance operational regimes relevant to the operational domain of MAST-Upgrade [2, 3], which is currently under construction.

Efficient plasma core heating relies on good confinement of the NBI fast ions (F.I.) so it is important to understand the transport mechanisms affecting fast ions in order to design and access high performance and advanced plasma scenarios. Collisional diffusion of fast ions is well understood [4] however, increasingly, anomalous transport processes are being recognised as a significant factor and is the motivation behind the present work in which a study is executed that examines in detail one of the mechanisms that causes redistribution of the fast-ion population within and loss of fast-ions from the plasma. In so doing, operational domains can be identified in which these redistribution and loss mechanisms are minimised and operation of the machine up to its full potential can be realised.

The phenomenon of core MHD affecting the fast-ion distribution in tokamaks has been known since the early 1980s [5] and has been observed and studied in more detail recently in MAST [6]. In the MAST device, typical plasma temperatures and beam injection energies result in neutron emission dominated by interaction of the beam fast ions with the background thermal plasma ions. Detailed observation of the neutron emission may therefore be used as a proxy measurement for the high energy part of the spectrum of the NBI F.I. distribution within the plasma. In the previous study [6], a drop in the neutron emission rate correlated with MHD activity was observed as NB fast-ions were expelled from the plasma or redistributed into regions in which the reaction rate was lower. It was further observed that reducing the fast ion pressure gradient with off-axis injection and power reduction was seen to reduce the fast ion redistribution. This inspired the present work to characterise the domain where such redistribution could be avoided and develop a means to extrapolate this to MAST-Upgrade.

## 2. Diagnostics and modelling

Included in the suite of high quality plasma diagnostics available on MAST are three devices which give complementary measurements that may be used to determine fast-ion behaviour in MAST. Firstly, a uranium Fission Chamber (FC) provides measurements of the total volume-integrated neutron emission rate. The FC is absolutely calibrated with time resolution of 10µs. Secondly, a radially scanning neutron camera (NC) [7] simultaneously views four chords through the plasma, two on the machine midplane with different tangency radii separated by 20cm and two angled downwards such that their tangency points are 20cm below the midplane with the same horizontal separation as the midplane chords. The impact parameter (i.e. the tangency radius of the line-of-sight of the viewing chord) of the camera may be changed between shots to achieve tangency radii in the range 0m < $R_{Tan,NC}$ <1.21m (positional accuracy of ±3cm), covering the core region of the plasma and beyond. This scanning capability, combined with the excellent repeatability of MAST plasmas, may be used to deduce profiles of neutron emissivity via forward modelling with fine spatial resolution. Each line-of-sight measures spatially collimated line-integrated 2.5MeV neutron flux from the plasma within each camera channel's field-of-view with a typical time-resolution of 1ms. Due to the finite acceptance angle of each chord, accurate modelling of the camera geometry is necessary to enable quantitative interpretation of the NC data.

Finally, a Fast-ion D-alpha (FIDA) system [8] provides further information about the F.I. population. The FIDA camera maybe set in either of two viewing configurations. The first is a near-toroidal view from a lens assembly mounted just above the vessel midplane with lines-of-sight intersecting the path of one of the neutral beams at major radii from 0.77m to 1.43m. The second view is near-vertical looking down onto the path of the neutral beam. 11 lines of sight are available which may be any mixture of toroidal or vertical views. Passive lines of sight with similar viewing geometries, but which do not intersect the neutral beams, provide background subtraction of bremsstrahlung and passive FIDA light (FIDA emission from edge neutrals).

The complete suite of high spatial and temporal resolution diagnostics on MAST allows accurate interpretive analysis to be carried out. These include a 160 point, 200Hz Thomson Scattering (TS) system [9] providing measurements of temperature and density. Edge neutral density and edge density gradients are provided by an absolutely calibrated linear D-alpha camera [10, 11]. Charge-exchange recombination spectroscopy provides measurements of ion temperature and plasma rotation profiles. A 2-D visible bremsstrahlung imaging camera [12], combined with the TS data, provides Z-effective measurements. A motional-stark-effect (MSE) diagnostic [13] provides a measure of the pitch angle of the magnetic field and can be used in conjunction with the EFIT code [14] to determine the plasma current profile. Together, these measurements provide a basis for interpretive modelling of the plasma.

For this study the interpretive code of choice is TRANSP [15]. TRANSP includes the NUBEAM Monte-Carlo neutral beam code which calculates the beam particle deposition profile and tracks the fast-particle guiding centres to thermalisation or loss from the plasma. A Larmor orbit radius correction to the guiding centre position is then used to determine the plasma parameters local to the particle position that are used in the collision operator. With a sufficiently large Monte-Carlo population, a fast-ion distribution function is created that is close to that which would be obtained from a full gyro-orbit calculation. A pre-processor code suite is used to self-consistently prepare the input data for the TRANSP simulations. The pre-processing stage ensures accurate mapping of necessary data to plasma equilibria derived from EFIT [14] using magnetic, kinetic and MSE constraints, which also provides the plasma boundary for the TRANSP simulations.

The two neutral beam injectors installed on MAST during the recent experimental campaign have identical geometry: horizontal injection at the machine midplane with tangency radius=70cm. The injectors themselves were identical except for the configuration of the ion sources. Ion sources on NB injectors produce a low density, low temperature plasma from which ions are extracted and accelerated to create the energetic particle beam [16, 17]. Neutrals with different energies are eventually created depending on the molecular composition and charge states that exist in the source plasma (principally $D^+$, $D_2^+$, $D_3^+$). Depending on the detailed configuration of the confining magnetic field on the ion source, the various charge species can be enhanced or suppressed with respect to one another. On one beamline the source was in a so-called "supercusp" configuration which enhances production of full-energy ($D^+$) component ions. On the other beamline the source was in a so-called "chequerboard" configuration which enhances production of half- and third-energy component ions. The different configurations (the names of which refer to the geometry of the magnetic filter on the ion source) were in use to assess their relative merits with a view to selecting one or the other as most beneficial for the forthcoming MAST-Upgrade. This results in the beams having different ratios of Full:Half:Third energy neutral components. For the results presented in this study the "supercusp" source provides power fractions in the ratio 0.89 : 0.08 : 0.03 and the "chequerboard" source in the ratio 0.56 : 0.27 : 0.17. This is not a problem in the analysis since the species mix is taken into account by the NUBEAM code.

The calculated total neutron emission rate from TRANSP may be compared with the global neutron emission rate as measured by the fission chamber diagnostic. TRANSP is also able to output the full F.I. distribution function $f(R, Z, E, v_{||}/v)$ (R,Z=spatial coordinates, E=particle energy, $v_{||}/v$ =pitch angle cosine with respect to the magnetic field, assumes toroidal rotational symmetry) as well as the neutron emission mapped on a 2-D poloidal cross-section at arbitrary times in the simulation. This data can then be processed with the code LINE2 [18] to produce profiles of neutron emission rate as measured by the scanning neutron camera. Thus measurements from two independent diagnostics may be compared with state-of-the-art interpretive modelling of experimental data. This combination allows robust conclusions to be drawn concerning the evolution of the F.I. population.

## 3. Experiments

### 3.1 Experimental basis and design

The growth rate (γ) of fishbone instabilities is related to gradients in the 6-D F.I. distribution function ($f_{FI}$) by [19, 20]:

$$\gamma \propto \omega \frac{\partial f_F}{\partial} - n \frac{\partial F}{\partial P_\varphi} \qquad (1)$$

where Š is the mode frequency, $E$ is fast particle energy and $P_\varphi = m\ v_\varphi - e$ (where m is the particle mass, R is the major radius, $v_\varphi$ is the toroidal velocity and $\psi$ is the poloidal flux function) is the canonical toroidal angular momentum. Using the approximation $P_\varphi = -e$ , $P_\varphi$ can be thought of as a negative radial co-ordinate so this relation implies that a negative energy gradient in the F.I. distribution function will decrease the instability growth rate whereas a negative radial gradient provides a source of free energy to drive the instability. In this experiment, the injection energy is kept constant so there should be little variation in the energy gradient between experiments. Changing the plasma density and/or the input power, however, should produce significant modification of the radial gradient of the distribution function. In increasing the power at fixed injection energy, the number of particles deposited at a particular location in the plasma will increase approximately in proportion to the increase in power at fixed plasma density. This increase in magnitude of the fast ion population will increase the magnitude of the spatial gradient (which is negative since the F.I. population tends to be peaked at the plasma magnetic axis) and increase the instability growth rate. A decrease of the plasma density allows deeper penetration of the beam into the plasma core and cause proportionally greater deposition of beam particles in the core. At lower density, an increase in temperature for given heating power will also be expected which will increase the slowing down time of the fast ions. These effects will both act to increase the F.I. pressure and hence increase the radial gradient of the F.I. distribution function.

The experiments carried out were therefore designed to explore a range of density and power levels in a plasma scenario that was operationally relevant to MAST-Upgrade. The intention of the experiment was to map out boundaries of the operational domain that could be accessed with low amplitude fishbone instabilities such that significant redistribution of NB fast ions is suppressed. The shot sequence carried out to explore the density/power parameter space comprised a 4-point scan in plasma density with 2 shots at each density level with 1 and 2-beam heating respectively, providing scans of density at 2 constant power levels or viewed the other way, scans of power at 4 constant density levels. Other plasma parameters such as position, shape, NB injection energy and NB start-time were kept as identical as possible shot-to-shot. In the 1-beam shots, the same beam (that fitted with the "supercusp" ion source) was used. At the injection energy chosen (60keV), the injected neutral power from this beamline was 1.5MW. Care was taken to obtain the same injection energy on each beamline ensuring that the full, half and third energy components had the same birth energies and steps in the energy spectrum were as close as possible to that present in the 1-beam shots. At the injection energy chosen, the neutral power injected by the second beamline (that fitted with the "chequerboard" source) was 1.6MW. Of course, considering the different ion sources in use on the two beamlines a small perturbation to the energy gradient is expected but this was assumed to be small compared to the changes effected in the $\partial\ /\partial P_\varphi$ term in equation (1). Motional Stark Effect (MSE) measurements may only be obtained when the "supercusp" beam is injecting power hence the choice of this for the single beam shots resulting in MSE measurements being available for all 8 shots in the parameter scan. At the highest and lowest density levels, the single beam shots were repeated a number of times with the scanning NC in a different position during each shot. This allowed the reconstruction of the neutron emission profile at the extremes of the density scan allowing further conclusions to be drawn concerning the spatial extent of the phenomena being studied.

The principal features of the plasma scenario used for this study were: a "direct-induction" start-up technique producing an 800kA flat-top plasma current (see Figure 1a), 0.585T toroidal field and neutral beams starting at 150ms (1-or 2-beam depending on the particular shot). The "direct induction" start-up was developed in the previous MAST experimental campaign to be as relevant as possible to MAST-Upgrade. The poloidal field coil set in MAST-Upgrade will not be compatible with merging compression, in which plasma is initiated around in-vessel poloidal field coils. Instead, "direct induction" forms the plasma in the machine mid-plane and use of this technique in the present study ensured that the early evolution of temperature, density and current profile was as close as possible to that expected in MAST-Upgrade. Figure 1b shows the time evolution of the line-integrated density for each of the four density levels studied. In the period of interest, between 250-270ms, these resulted in line-averaged density levels of between 2.0 and $4.5\times10^{19}m^{-3}$. The radial profiles of the electron density as measured by the Thomson scattering diagnostic at 260ms are shown in figures 1d-g demonstrating very similar density profiles for the 1-beam and 2-beam shots at each density level.

At the temperatures typically reached in MAST (typically 800eV< $T_e$ <1500eV) fusion-neutron production is dominated by beam-thermal and beam-beam interactions which account for >99% of the total emission. Thermonuclear reactions contribute a negligible proportion of the total (<<1%). This makes the neutron emission profile a useful diagnostic tool to track changes in the fast-ion distribution and any observed drops in the neutron emission rate can be interpreted as a reduction of the F.I. reaction rate in the region of the plasma under observation. In the absence of any other changes to plasma parameters, this can in turn be interpreted as a loss of fast-ions from that region or, in the case of a global measurement, a redistribution of the fast ions into regions at lower temperature where the steady-state reaction rate will be reduced [21] and/or ejection of fast-ions from the plasma.

Figure 2 details the MHD activity from the 8 shots comprising the power/density scan and shows, for one representative shot, the correlation between the MHD activity and abrupt changes in the neutron emission rate. Figure 2(a-h) show time traces obtained using a Mirnov coil mounted on the Low-Field-Side of the vessel (LFS-MC) for the 8 shots in the power/density scan. It is obvious from simple inspection of these time traces that higher power at constant density or lower density at constant power leads to higher amplitude MHD modes. Figure 2i shows the spectrogram of the low density 2-beam shot (29221, figure 2e) representing the case with the highest amplitude MHD activity. The "chirping" modes that are of particular interest in this investigation are seen to occur from about 220ms onwards in this shot. Figure 2j shows the neutron emission rate from the low density 2-beam shot obtained using the fission chamber and the two midplane chords of the scanning neutron camera. The purple line shows the RMS amplitude of the Mirnov coil signal. Simple inspection of these signals show that the abrupt drops in neutron emission rate measurements are well-correlated with the MHD modes. The high density 1-beam shot representing the case with completely suppressed MHD activity (shot 29195) has also been used as the reference case in [22] in which results are presented concerning the modelling of different perturbations to classical fast ion simulations and their comparison with neutron camera measurements.

From the spectrogram (figure 2i), it is clear that the beam-on period can be broadly divided into 3 distinct phases. In the early phase from beam-on at 150ms to around 220-250ms broadband toroidal Alfvén eigenmodes are apparent with a frequency in the range 50-100kHz. Next, from 250 to around 280-300ms, there is a phase which contains principally "chirping" modes in the frequency range 20-50kHz or, in the high density 1-beam case, is practically MHD free. From 280-300ms, the so-called long-lived mode (LLM), associated with the existence of a q=1 surface in the plasma, appears and continues to the end of the shot. These phases occur at similar times in all 8 shots in the parameter scan as the q-profile evolves throughout each shot. In MAST-Upgrade it is desirable to develop an operational scenario with q just above 1 in the core of the plasma to avoid the long-lived mode (which can be seen in the MHD spectrogram for MAST shot 29221 in figure 2i starting at ~290ms). It is therefore the phase with the core chirping modes that is of interest as the plasma in this phase, particularly the form of the q-profile, most closely resembles that expected to be favourable for such stationary operation of MAST-Upgrade. Analysis has therefore focussed on a short interval just before the appearance of the long-lived mode and is referred to as the "period-of-interest" (POI). This is the period 250-270ms for all the shots with one exception. In shot 29251 (medium-high density, 2-beams), a large amplitude low-frequency mode (identified as a neoclassical tearing mode) is apparent that starts at around 200ms and survives to 260ms. Since the presence of this mode may also influence the F.I. distribution and is not specifically the type of MHD activity that is of interest, the POI for this shot is taken as 260-280ms. Some residual effect of the low-frequency mode can be expected during the period 260-280 but, since the MHD activity during this period appears to be very similar to that used in the other shots and the appearance of the LLM occurs after 280ms, the period 260-280ms for this shot is considered more representative of the effect of the core chirping modes. The POI is indicated by vertical dashed lines in all panels of Figure 2. Defining these POIs allows time averages of various quantities to be taken during a well-defined period for each shot, illustrating the effect of intermittent bursting modes on the F.I. population and therefore the resulting plasma heating and current drive.

## 3.2 Comparison of results with Modelling

As discussed in section 2, the TRANSP code was used in conjunction with the high quality data available from the MAST diagnostics to produce interpretive simulations of the 8 shots in the power/density scan. A figure of merit that can be examined to assess the quality of a TRANSP run is the calculated neutron emission rate as compared to the rate measured by the fission chamber. Of course, for most of these shots, the neutron emission rate is not a good measure as this is strongly affected by MHD. Hence examination of the calculated and measured neutron rate for the high density 1-beam case (shot 29195, figure 3a (inset)) shows that, in the absence of MHD, the neutron rate calculated by TRANSP is in excellent agreement with the rate measured by the FC. Furthermore, a detailed sensitivity study by Klimek and co-workers [22] has also shown that typical uncertainties in the experimental data used in the interpretive modelling result in much smaller uncertainties in the neutron emission rate than the changes observed during MHD activity. We conclude therefore that such changes arise from genuine effects of MHD on the F.I. population rather than being due to experimental errors.

When performing the TRANSP simulations, no information was included in the F.I. transport model concerning the effect of the MHD, hence any mismatch between the measured neutron emission rate and the calculated neutron emission rate can be assumed to be due to the MHD affecting the F.I population. Examination of the graphs in the inset of Figure 3a show that as the density is reduced (bottom to top) and/or the power increased (left to right), the mismatch between the measurement (solid lines) and the TRANSP calculation (dashed lines) increases indicating a stronger effect on the F.I. population at higher power and/or lower density. Averages of the measured and modelled neutron rates during the POIs were calculated, and the ratios are shown in figure 3a (main) plotted versus line-averaged density. The error bars arise from combining the standard deviations of the measured and modelled neutron rates during the POI. A decrease in the ratio with decreasing density and increasing power is evident from this graph. Furthermore, the points representing the 1 beam shots at the two higher density levels have ratios equal to 1 within the uncertainty, indicating that in these two cases the F.I. population was essentially unaffected by MHD. The ratio for the 2-beam shot at the highest density also has an error bar that crosses unity although this is at the very extremity of the error range so it is likely that some F.I. redistribution did occur in this shot. The two higher density 1-beam shots are therefore taken as representing the boundary in the density/power domain consistent with operation with no (or very little) F.I. redistribution.

From the TRANSP simulations, it is possible to obtain a poloidal projection of the neutron emissivity. This projection is assumed to be toroidally symmetric but, unlike most TRANSP outputs, is not flux-surface averaged, a point that is important when interpreting the results from the neutron camera (NC) [23]. The LINE2 code [18] uses this neutron emissivity projection to model the full profile of the expected neutron emission rate measured by the NC in any particular position. The amplitude of the computed profile is the only free parameter in the calculation and can be thought of as the NC's detection efficiency. In practice, the detector efficiency for all lines of sight has been independently determined as being around $0.08 \pm 0.02$ [7] and the "free" parameter in the LINE2 calculation is expected to be close to this figure (in the absence of MHD influences on the F.I. population). NC profiles have been calculated in this way for the 8 shots of the power/density scan using the outputs from the TRANSP simulations. In the experiments performed, only the highest density single beam discharge (29195) was determined to be free of MHD and redistribution effects during the POI (as demonstrated in figure 3a). Therefore, this shot was used to determine the efficiency factor of the NC lines of sight such that measurements from all 8 shots may be compared with the calculations in a consistent manner. The detector efficiency factors calculated in this way were found to be 0.08 for both midplane viewing chords.

Figure 3 (b-c) shows the results of this exercise. Once again, the comparison between the measured and calculated count rate is averaged over the POI and expressed as the ratio (measured rate) /(calculated rate). The error bars indicate the combination of the uncertainty in the NC count rate inherent to the detectors and the standard deviation of the average over the POI. These sources of error are additive as the first is due to the characteristics of the instrument and the second is due to temporal variations in the plasma. In all 8 shots, the NC was in the same position such that its midplane chords had impact parameters of 0.948m and 1.14m. These will be referred to in the following discussion as on-axis and off-axis respectively. It is finally noted that the off-axis measurement for shot 29034 (mid-low density, 1-beam) was determined to be faulty so has been omitted from this particular analysis.

In agreement with the FC results, the ratio for the on-axis chord (figure 3(b)) decreases with decreasing density and the values of the ratio are similar to the FC results at each density level. The error ranges are of a magnitude such that it cannot be said with any certainty that there is an apparent trend for decreasing ratio with increasing power though such a trend is not precluded by this analysis. The neutron emission ratio for the off-axis chord (figure 3(c)) also exhibits a decrease in the ratio at low density though at the mid-high density level (~$3.5 \times 10^{19}$ m$^{-3}$) the ratio for the off-axis viewing chord is above unity for both power levels.

A possible explanation is that, at this density level, the F.I. redistribution effect is acting in such a way as to increase the density of fast particles at the radius observed by this chord. Unfortunately, further investigation of this observation is not possible with the data set available.

The TRANSP code also includes the facility to output the full fast-ion distribution function (FIDfn) at specified times. Using this distribution together with the standard TRANSP output, it has been possible to calculate the canonical toroidal angular momentum of particles in the F.I. distribution and hence calculate $\partial /\partial P_\varphi$ i.e. the term in equation 1 driving the MHD modes. In each case the largest absolute value of $\partial /\partial P_\varphi$ over all energy, pitch-angle and $P_\varphi$ is taken as representative of the magnitude of the FB driving term. No attempt has been made in this instance to identify the particular resonance giving rise to the MHD modes observed, hence the absolute values of $\partial /\partial P_\varphi$ are only presented here to provide a comparison between the maximum possible value of the driving term for the MHD in each case. The results of this calculation are shown in figure 4(a). The calculation of $\partial /\partial P_\varphi$ was performed in the same way for each of the TRANSP simulations in this study, hence the results are shown on the same arbitrary scale and may all be directly compared. For the shots under investigation, the FIDfn was output at the start, middle and end of the POI enabling an estimate of the error in the driving term due to temporal variations in the FIDfn, which is the basis of the displayed error bars. Inspection of figure 4a clearly shows that the magnitude of the MHD driving term decreases with increasing density and also decreases with decreasing NBI power, as expected. In an earlier study by Turnyanskiy et al [6], a 1-beam/2-beam pair of shots at constant density was obtained. The original analysis of these shots used an ad hoc anomalous fast ion diffusion term ($D_{an}$) to characterise the magnitude of the F.I. redistribution in the modelling which indicated a low-level of F.I. redistribution for the 1-beam shot (shot number 26887: $D_{an}$=0-1m$^2$s$^{-1}$) and much higher level for the 2-beam shot (shot number 26864: $D_{an}$=2-3m$^2$s$^{-1}$). This pair of shots has been re-analysed to obtain the fast-ion distribution functions from NUBEAM such that comparison can be made with the 8 shots of the present study. The results for these two shots are also shown in figure 4(a) and are consistent with the present study i.e. the absolute value of $\partial /\partial P_\varphi$ for both shots in the previous study is comparable with that calculated for the shots in the present study. Also, the value of $\partial /\partial P_\varphi$ for the two beam shot is significantly larger than for the one beam shot, in agreement with the present study.

The relative magnitudes of the MHD drive term can now be compared with the observed growth rate of the MHD in each case. As part of the standard data, the MAST magnetics team provide an estimate for the absolute value of the magnetic perturbation due to MHD modes in the plasma. The derivative of this signal was therefore used to estimate the maximum value of the mode growth rates for all modes that occurred during the POI in each shot. In those shots in which more than one mode occurred during the POI, an average growth rate has been calculated. The results are plotted in Figure 4(b) versus the line averaged density and are grouped into 1-beam and 2-beam shots. No modes occurred during the POI for the highest density 1-beam shot (29195) so this data point has been set to zero. The error bars for the growth rate represent the standard deviation of the calculated growth rate in a 50μs-wide window centred on the point at which the growth rate of the mode reaches its maximum value. The error bars for the density data represent the standard deviation of the line-averaged density during the POI. The trends in the mode growth rate are seen to be consistent with the $\partial /\partial P_\psi$ estimates, i.e. that higher power and/or lower density shots which exhibit higher values of $\partial /\partial P_\psi$ also exhibit larger mode growth rates. It is noted that the mode growth rate for the lowest density two beam shot shows a slightly lower value than expected. The estimated error bounds on $\partial /\partial P_\psi$ however mean that this point is not inconsistent with the gradients in the calculated F.I. distribution. Data points from the previous experiments [6] are again included on this plot and are seen to be generally consistent with the present study; the growth rate in the two beam shot is significantly larger than in the one beam shot and both are of comparable magnitude to the results from the present study.

These modelling results taken together provide a coherent explanation of the relationship between the magnitude of the F.I. redistribution observed and the position of the data point in the power/density parameter space. Essentially, lower plasma density and/or higher power leads to a larger value of the MHD drive term in equation 1. This is followed by a generally higher observed value of the mode growth rate which is then followed by a larger magnitude redistribution effect as determined by the deficit in the measured neutron production rate compared with the modelled neutron production rate.

### 3.3 Empirical evidence for core localisation of F.I. redistribution effect

**Neutron Camera**

The chirping modes in the POI are identified as n=1 "fishbone" modes [6] where n is the toroidal mode number. These modes are thought to be localised to the plasma core and ascertaining the radial extent of their influence on the F.I. population is one of the subjects of this investigation. Since both the NC and FIDA systems have good spatial resolution it has been possible to determine the spatial extent of the F.I. redistribution effect of the FB modes using these diagnostic data.

NC position scans were performed in a series of repeated shots at the highest and lowest density levels. In the low density case, 5 shots were performed, which were nominally identical apart from the viewing angle of the NC, giving a total of 10 spatially separated chords on the midplane. In the high density case, 6 shots were performed giving 12 chords. In the high density scan, small fishbone modes were observed during the POI in 3 of the 6 shots. However, the effect on neutron emission rate at the time of these modes was comparable to the noise level in the measurement hence no statistically significant result could be determined and the high density case was deemed free of F.I. redistribution effects. In the low density case, however, the neutron emission rate was somewhat higher and the MHD modes were of larger amplitude producing drops in the NC count rate much larger than the noise. The fishbone modes occurred during similar time intervals in all 5 shots of the low density scan so detailed analysis of the NC data from the various lines-of-sight over the POI has been carried out to determine the spatial effect of the FB mode on the F.I. population.

For each viewing chord the analysis method illustrated in figure 5 was followed. Firstly, the onset time of each FB mode during the POI was determined by setting a threshold in the amplitude of the RMS of the LFS-MC signal; the time at which the signal reached this amplitude was labelled $t_{thresh}$. Examination of the LFS-MC signal determined that the FB modes typically started approximately 0.5ms before the modes reached this amplitude and were typically of 2.5ms duration. This period, $(t_{thresh}-0.5ms) < t < (t_{thresh}+2.0ms)$ was taken to be representative of the period during which the FB mode was affecting the F.I. population. These intra-FB periods are highlighted in green in figure 5. The start and end times of inter-FB periods (i.e. the periods between $FB_i$ and $FB_{i+1}$ in which no redistribution effect was occurring) were defined by $(t_{thresh,i}+5ms) < t < (t_{thresh,i+1}-0.5ms)$ i.e. 5ms after the start of one fishbone to the start of the next. The start and end time of the POI make the final boundaries of the inter-FB assessment periods (unless the start/end times happen to be defined as being within intra-FB periods). These periods are illustrated by red lines in the figure and the average over these periods, taken to represent the unperturbed neutron emission rate, is referred to below as $NC_{AV}$.

The drop in neutron rate due to the effect of a particular FB mode was then determined by examination of the signal during the intra-FB periods as:

$$\Delta N_{FB} = \max(N_{i,-F}) - \min(N_{i,-F}) - \left(\sqrt{2N_S^2}\right)$$

where the final term is the combination in quadrature of the statistical uncertainty in the neutron count rate at the maximum level and at the minimum level. Inclusion of this final term accounts for the uncertainty in the magnitude of the neutron emission signal ensuring that the final value for $\Delta NC_{FB}$ is a reasonable evaluation of the drop in the signal due to the FB and not simply an artefact of the noise in the signal. The statistical uncertainty was assessed using the inter-FB periods in the time interval 250-270ms in shot 29222 and was found to be 10.5% of the signal level. The value of the statistical uncertainty was therefore set to $NC_{std}=0.105\times\max(NC_{intra\text{-}FB})$. The value of $\Delta NC_{FB}$ is indicated in figure 5 for the first of the FB modes in this POI in which the subtraction of the final term in the above expression is indicated by the vertical blue lines above and below the resultant $\Delta NC_{FB}$. For shots in which more than one FB event occurred in the POI, the average value of $\Delta NC_{FB}$ for all events was taken as a representative single value for that NC chord. For those FB events in which the error exceeded the difference between the maximum and minimum emission rate, i.e. in which $\Delta NC_{FB} \leq 0$, the value was set to zero.

The proportional drop in neutron emission was then finally computed as $\Delta NC_{FB}/NC_{AV}$ for each of the NC chords. The error in the final value was taken to be $\sqrt{2N_S^2 + N_{A,S}}$ where the first term represents the error in determining the magnitude of the drop in the neutron rate and the final term the error in determining the average of the inter-FB signal.

Ideally, the FB modes would be spaced sufficiently that the neutron emission rate could return to its completely unperturbed level before the next FB mode again affects the F.I. population. In the example shown, it is apparent that this is not the case so the estimate of signal noise, $NC_{AV,std}$, is probably an overestimate since it necessarily includes some of the recovery period of the F.I. population. This overestimate of the error actually serves to strengthen the following conclusions since, even with the larger error bars, the drops in the neutron emission rate are still large enough to provide evidence of the localisation of the effect on the F.I. population.

The result of this analysis is shown in figure 6a where the black data points are $\Delta NC_{FB}/NC_{AV}$ as a function of the impact parameter of the NC chords. Also shown is the absolute neutron emission rate (red data points, plotted on an arbitrary scale) to illustrate that the proportional drop in neutron emission is not simply a function of the absolute neutron emission rate. Various plasma equilibrium quantities have been taken from the MSE-constrained EFIT and are displayed on the graph along with the NC data. The midplane radius values of the plasma boundary are indicated by vertical dashed black lines and the radial position of the magnetic axis is shown by a vertical dashed green line (the standard deviation of the variation in position of the magnetic axis throughout the POI is indicated by the grey region). The radial position of the q=1.5 surface is also shown. It is clear therefore that all NC chords examined have tangency radii well within the q=1.5 surface. No q=1 surface was determined by MSE-constrained EFIT for these shots, however n=1 fishbones are associated with a q=1 surface. It is therefore probable that the q profile drops to a value close to, or equal to 1 in these plasmas during the POI and this is consistent with the typical error in the EFIT derived q-profile.

Of particular note in this graph is the large localised peak at 1.0m < R < 1.1m indicating that the F.B mode is having a proportionally larger effect on the F.I. population in this region as compared with the rest of the profile. This is therefore interpreted as evidence that the effect of the FB mode is localised in the plasma to a defined region just outside the magnetic axis. The observation that the magnitude of $\Delta NC_{FB}/NC_{AV}$ has a local minimum at the magnetic axis where the neutron emission rate itself has a local maximum, as well as the fact that $\Delta NC_{FB}/NC_{AV}$ decreases with radius more rapidly than the neutron emission rate, also leads to the conclusion that the peak in $\Delta NC_{FB}/NC_{AV}$ is not simply a consequence of a lower level of neutron emission at larger radii but is genuinely due to localisation of the effect of the FB mode.

Interestingly, there is a second local maximum in the $\Delta NC_{FB}/NC_{AV}$ data on the high field side (HFS) of the magnetic axis at R~0.75m. Examination of the full magnetic flux reconstruction from EFIT indicates that the maxima are located at similar values of normalised poloidal flux (the position of $\rho = 0.07$ is shown by a blue vertical dashed line where $\rho$=normalised poloidal flux). The most likely explanation for this profile concerns the geometry of the lines-of-sight of the NC. If a line-of-sight has a tangency radius on the low field side (LFS) of the magnetic axis then it only sees the LFS of the plasma. If the tangency point is inboard of the magnetic axis, it looks through the LFS as well as at the HFS. Measurements from such lines-of-sight will therefore show artefacts associated with activity on the LFS. However, since the line-of-sight has a shorter path length in the LFS of the plasma compared to a chord with larger tangency radius, any line-integrated measurement due to LFS activity will be of lower amplitude. Since the HFS peak in figure 3 is of lower amplitude, it is certainly possible that this is the explanation for the presence of this peak, i.e. a relatively steady but lower amplitude neutron emission rate from the HFS compounded with the large drops in neutron emission rate from activity on the LFS.

The detailed analysis presented here of the low density 1-beam case can be compared with the 2-point analysis shown in 3(b-c) for the other power and density levels in the scan. First it is noted that the points in figure 3(b-c) representing the 1-beam low density shot are consistent with the analysis shown in figure 6a. Even though the 0.95m line of sight corresponds to a local minimum in figure 6a, this is still consistent with the error range of the ratio for this point in figure 3b. It is noted that a line of sight with major radius 1.05m corresponding to the position of the local maximum in figure 6a, would produce a data point in figure 3b with ratio ~0.58 indicating a larger effect on the F.I. population at this radius than at 0.95m. Data for radius 1.05m is only available for the low power/low density point in the scan so it is not possible to produce equivalent graphs to figures 3(b-c) for any other radii.

**Fast Ion D⌐**

The FIDA diagnostic recorded data for the 8 shots of the power/density scan using core orientated toroidal lines-of-sight. The FIDA data have been processed to obtain the proportional drop in the FIDA signal due to the FB modes in the same way as the NC data described above. The FIDA signal suffered from relatively higher noise than the NC and the signal intensity decreased markedly with increasing plasma density, as is expected from the physics of FIDA spectroscopy. The FIDA data at the two highest density levels were therefore unusable, exhibiting unacceptably low signal to noise ratio but the data from the four plasma shots at the two lower density levels contain useful information. The analysis of the 1 and 2-beam shots at the 2 lower density levels is shown in Figure 6(b-e). In these graphs, the black data points indicate the proportional drop in the FIDA signal due to FBs during the POI, plotted versus the impact parameter of the FIDA lines-of-sight. The red data points show the absolute value of the FIDA signal (plotted on an arbitrary vertical scale) as a function of the impact parameter. The error bars represent the standard deviation in the signal over the POI. The vertical green line indicates the position of the magnetic axis as determined by MSE constrained EFIT (the grey region indicates the standard deviation of the position over the POI).

Comparison of the absolute FIDA signal and the proportional drop once again indicates that the drop in the signal due to FBs is not simply a function of the magnitude of the signal. Even in the case of the lowest density 1-beam shot where this could be a valid interpretation at R<1.1m, the lines diverge rapidly for values of R>1.1m. Although the errors in the proportional-drop analysis are generally larger for FIDA than for the NC, the conclusion can still be drawn that F.I. redistribution due to FBs has a greater effect close to the magnetic axis, i.e. in the core of the plasma, than at larger radii (R>1.1m) even though a substantial F.I. population exists at larger radii as indicated by the magnitude of the FIDA emission at

R>1.1m. This analysis is therefore in qualitative agreement with the NC analysis in the previous section and provides further evidence for the core localisation of the effect of FB modes on the F.I. population using a diagnostic technique that is completely independent of the neutron emission analysis discussed previously

## 4 Implications for MAST-Upgrade scenarios

A significant upgrade to MAST is currently under construction [2, 3] that will bring increased capabilities to the experiment, such as off-axis NBI heating and current drive, increased toroidal field and increased pulse duration as well as testing the novel Super-X divertor concept. The off-axis NBI current drive (NBCD) is intended to allow a certain degree of control of the q-profile with the intention of creating plasma scenarios with current profiles tailored to avoid deleterious MHD modes. In the Core-Scope (CS) upgrade, due to be operational in 2015, the NBI system will consist of two PINI injectors, one located with the same injection geometry as those currently installed on MAST (referred to hereafter as the "on-axis" position), the other located with a horizontal line-of-sight 650mm above the midplane (referred to hereafter as the "off-axis" position). This combination will provide efficient core heating, via the on-axis PINI, and significant off-axis heating and NBCD via the off-axis PINI.

The maximum specifications of these injectors are injection energy up to 75keV and power up to 2.5MW. In the experiments described earlier in this paper, the 2-beam shots had on-axis injected power of around 3MW, a similar level to the maximum achievable on-axis power in the MAST-Upgrade Core Scope. Following from the experimental results presented in this paper, the implication of operation of the on-axis MAST-Upgrade beam at full power is that F.I. redistribution via FB modes is likely to occur at all but the highest density levels. Since one of the purposes of the NBI configuration on MAST-Upgrade is to allow q-profile tailoring, uncontrolled redistribution of the F.I. population by MHD would then limit the degree of control that can be effected by the re-positioned NB injector.

Development of steady-state plasmas, in which the plasma current is maintained by fully non-inductive current drive for a period of several current diffusion times, will be important for future spherical tokamak applications. A significant proportion of this non-inductive current is intended to be provided by NBCD which will require operation at low density. As the results in this study have shown, operation at low density is incompatible with avoidance of MHD and redistribution of the F.I. population. However, these results also suggest a strategy for modification of the NBI system that will be installed in later stages of the upgrade that should produce significant mitigation of the MHD responsible for the observed F.I. redistribution.

In the staged construction plan for MAST-Upgrade, the principal modification in moving beyond Core Scope will be the addition of a new beamline called the double beam box (DBB) which will give the capacity for an extra on-axis beam at the midplane and/or an extra off-axis beam 650mm above the midplane. Throughout the design of MAST-Upgrade, physics assessments of the expected performance of the plasma using different assumptions of engineering geometry and capability have been carried out using TRANSP. The TRANSP simulations were set up using carefully selected temperature and density profiles that are expected to be achieved in MAST-Upgrade together with realistic plasma shapes that can be produced with the MAST-Upgrade poloidal field coil set [2, 3, 24]. Two representative plasma scenarios are used with Greenwald density levels $n_0/n_{GW}$=0.58 and $n_0/n_{GW}$=0.23, identified as scenarios A1 and A2 respectively, where $n_G = I_p/\pi a^2$ ($10^2 /m^3$), $I_p$ is the plasma current in MA and $a$ is the plasma minor radius in metres and $n_0$ is the density at the magnetic axis. The high density scenario, A1, is used to assess performance in a standard H-mode whilst the low-density scenario is used to assess maximum expected current-drive capabilities of the NB system and potential NB shine-through issues. In all simulations, the electron temperature was adjusted in order to achieve an H-factor of 1 ± 0.03 according to the IPB98(y,2) H-mode scaling [25]. The MAST-Upgrade simulations differ from the experiments discussed in section 3 in that they are carried out with a plasma current of 1MA and $B_0$=0.785T compared to 0.8MA and $B_0$=0.585T, however the plasma current does not significantly affect the fishbone stability calculation [19] and the ratio $I_p/B_0$ in the two cases is similar, so the comparisons that will be shown in this section between the MAST-Upgrade TRANSP simulations and the MAST experiments is still considered valid.

It has been shown in the previous sections that the experimental results are consistent with the FB stability theory. Therefore, the principal factor determining whether F.I. redistribution via fishbones is likely to occur has been demonstrated to be the gradient of the F.I. distribution with respect to canonical toroidal angular momentum. With the representative TRANSP runs available for MAST-Upgrade, it is possible to examine the resultant F.I. distribution functions and assess them in comparison to the experiments. The results of this study are shown in figure 7. This diagram is essentially the same as that shown earlier in figure 4a with additional data points representing the results from the assessment of various MAST-Upgrade plasma scenarios and NBI injector combinations and geometries.

The two highest density 1-beam shots in the MAST experiments were shown to have neutron emission rates consistent with very little, or no, F.I. redistribution. It can be stated therefore that any plasma scenario with $\partial / \partial P_\varphi$ at a similar level as these two shots is likely to be free of F.I. redistribution effects. Furthermore, the highest density 2-beam shot was marginal, having an error bound that included unity in figure 3a and the one beam shot from the previous experiments [6] exhibited a low-level of F.I. redistribution. Plasma scenarios with $\partial / \partial P_\varphi$ higher than this level are therefore expected to exhibit F.I. redistribution effects with the effect getting stronger as the gradient term increases in magnitude.

Various symbols have been placed on the graph representing the analysis of the MAST-Upgrade TRANSP simulations. The filled green triangles show the calculated maximum absolute value for $\partial / \partial P_\varphi$ of the Core Scope scenarios A1 and A2 (recall that in the Core Scope there will be one on-axis beam and one off-axis beam). Clearly, both of these points lie in the region in which the F.I. population should be unaffected by F.I. redistribution. The filled red triangles indicate the expected F.I. gradient in a future upgrade design (i.e. the Core Scope arrangement with an extra off-axis injector). In this case the magnitude of $\partial / \partial P_\varphi$ has increased in both the low and high density scenarios and the point representing the low density scenario is now in the region at which a low level of F.I. redistribution can be expected. The red triangle at low-density is actually an underestimate since, in order to allow the TRANSP simulation to run successfully, it was necessary to reduce the power of the on-axis injector to 1.5MW (c.f. 2.5MW in the other MAST-Upgrade simulations). This point should therefore have a higher value of $\partial/\partial P_\varphi$ and correspondingly larger amplitude MHD and F.I. redistribution.

In reality, it may well be the case that higher on-axis power will cause higher density peaking than has been assumed. The electron density profiles used in the MAST-Upgrade simulations are extremely flat and the difference between the low and high density cases is a simple scaling factor, i.e. no change to the density peaking factor has been included. This is a fair assumption since MAST-Upgrade will operate in H-mode in which flat density profiles can be expected. The flat density profile in these simulations, compared with the peaked density profiles seen in figure 1d, is one of the principal reasons for the low value of $\partial/\partial P_\varphi$ for the low density Core Scope MAST-Upgrade plasma. The possibility of density peaking occurring still exists though, particularly in the low density case in which the beams will fuel the plasma core more efficiently.

To mitigate F.I. redistribution in future upgrades beyond Core Scope, it is proposed that the on-axis beam in the DBB be tilted upwards from the horizontal such that the tangency point of its line-of-sight through the plasma is significantly above the midplane (this distance is referred to as $Z_{RT}$). The intention is to deposit more fast particles in a moderately off-axis location, contrasting with the core deposition provided by the on-axis beam and the off-axis deposition provided by the off-axis beam. This should result in a F.I. population which is more evenly distributed over the plasma radius, thus reducing $\partial/\partial P_\varphi$ in the F.I. distribution and the resultant drive for the deleterious MHD.

A number of simulations were carried out covering a range of inclination angles for the on-axis DBB NB injector, each geometry defined by the height of the tangency point of the line-of-sight above the mid-plane ($Z_{RT}$). The maximum inclination investigated corresponded to $Z_{RT}=300$mm which results in an inclination of the beamline of 12º upwards from the horizontal. A range of density levels were also investigated from the Greenwald density limit (approx. twice the Scenario A1 density) down to the lowest level that could be successfully simulated by TRANSP with this level of injected beam power (about twice the Scenario A2 density). For the $Z_{RT}=300$mm case an additional simulation was executed at the scenario A2 density level with reduced power in the on-axis beam to compare the effect of the DBB modification with the equivalent scenario A2 run (with two off-axis beams at 2.5MW and one on-axis beam at 1.5MW, shown as the red triangle at low density in the main plot of figure 7). In each case, the maximum value of the $\partial/\partial P_\varphi$ has been assessed in the same way as described in section 3.2.

The dashed lines in figure 7 (main) show the results of the MAST-Upgrade simulations including the inclined beam (i.e. one on-axis beam, one off-axis-beam and one tilted-beam with total 7.5MW injected). At the density levels achievable in the simulations, these cover a relatively small range of $\partial /\partial P_\varphi$. For this reason an inset graph has been provided in figure 7, showing a larger scale plot of the grey area in the main plot. Each line represents a different degree of inclination of the beam, identified by the parameter $Z_{RT}$ which runs in 50mm steps from 100mm to 300mm (the extrema of the scan are labelled in the inset plot). The main point to note from this graph is that the difference in $\partial /\partial P_\varphi$ at a density of $\sim 4\times 10^{19}$m$^{-3}$ is more than a factor of 2 between the highest and lowest beam inclination angle. The extra low density point for the largest beam inclination angle ($Z_{RT}$=300mm), when compared with the higher power simulation indicated by the red triangle at this density level shows that the modified geometry has had a marked effect on the value of $\partial /\partial P_\varphi$ reducing it to a level comparable with the 2-beam Core Scope simulation (indicated by the green triangle).

These simulation results demonstrate that the modification of the NBI geometry should be extremely beneficial in helping to reduce or avoid the effect of redistribution of fast ions by MHD in MAST-Upgrade. Assessments by the MAST-Upgrade engineering team indicate that tilt angles of the DBB beam close to the maximum proposed here should be achievable without major re-design of the DBB itself. With the proposed modification incorporated into the design, more control over the precise deposition location of beam fast ions is made available to the experimentalist, allowing effective heating and current drive as well as control for fast ion experiments.

## 5 Conclusions

Results have been presented from 8 plasma shots taken on MAST during a recent experimental campaign comprising a systematic 4-point scan in density and a simultaneous 2-point scan in injected neutral beam power in plasmas designed to be as similar as possible, the only major difference being the plasma temperature. Interpretive modelling of these plasmas demonstrates that operation at lower density and/or higher injected power leads to an increase in the magnitude of the gradient of the fast ion (F.I.) distribution function with respect to the canonical toroidal angular momentum, identified as the driving term for fishbone MHD modes. It has further been demonstrated that plasmas with higher absolute magnitude of $\partial /\partial P_\varphi$ exhibit higher amplitude n=1 fishbone MHD modes at similar times in the plasma evolution in which the minimum value of q in the plasma is close to 1. Comparison of total neutron emission rate, as measured by a fission chamber, with the predicted neutron emission rate modelled by the TRANSP code shows good agreement in MHD quiescent plasma with an apparent neutron deficit in the measurement in plasmas with larger amplitude MHD activity. The timing of abrupt drops in the neutron emission rate is well-correlated with the appearance times of the fishbone modes (in agreement with previous studies [5, 6]) so the apparent neutron deficit in the measurement is attributed to redistribution of the F.I. population by the MHD. It has also been demonstrated that the magnitude of the redistribution effect is related to the magnitude of the fishbone growth rate which, in turn, is related to the magnitude of the F.I. radial pressure gradient through equation 1. Significantly, a region of the parameter space investigated at high density and relatively low beam power has been demonstrated to be free of F.I. redistribution by MHD either through the complete absence of fishbone modes or due to the low growth rates of fishbones leaving the F.I. population relatively unperturbed.

Results from the scanning neutron camera have provided a detailed profile of neutron emission in a plasma exhibiting moderate levels of MHD activity and F.I. redistribution effects. Analysis of this profile has provided good evidence for localisation of the effect of the fishbone modes to a particular position in the plasma core but at an off-axis location. Similar analysis of the FIDA results produce a result that is consistent with the NC result, confirming that the effect of the fishbone on the F.I. population is localised to the core of the plasma. Results obtained using the three diagnostics on MAST sensitive to the fast ion distribution thus provide a consistent picture of MHD-induced fast ion redistribution in the plasma.

The results obtained during this study have been used in conjunction with detailed modelling of the expected performance of MAST-Upgrade. It has been demonstrated that F.I. redistribution by MHD should not occur during the core-scope stage of MAST-Upgrade, even at full beam power. It has further been shown that the addition of extra NBI power in future upgrades beyond Core Scope may produce moderate to high levels of F.I. redistribution, with commensurate losses of q-profile control and NB current-drive. However a small modification to the geometry of the on-axis DBB beam in future stages of MAST-Upgrade should produce a significant reduction in the magnitude of $\partial / \partial P_\varphi$ in the plasma, reducing the drive for MHD and the subsequent redistribution of the fast ions.

The results presented in this study have been shown to agree qualitatively with the theory of fishbone stability. Full quantitative analysis will require the coupling of predictive transport codes such as pTRANSP or JINTRAC [26, 27] with an MHD stability code such as MISHKA [28] and a wave-particle interaction code such as HAGIS [29]. Work is ongoing to make such code suites available and it is intended that re-analysis of the plasmas from the present study will be undertaken when the required tools become available.

## Acknowledgments

The authors are grateful to Eric Fredrickson for participation in the experiments and subsequent discussions and to Sergei Sharapov for discussions concerning fishbone stability. This project has received funding from the European Union's Horizon 2020 research and innovation programme under grant agreement number 633053 and from the RCUK Energy Programme under grant EP/I501045. To obtain further information on the data and models underlying this paper, please contact PublicationsManager@ccfe.ac.uk. The views and opinions expressed herein do not necessarily reflect those of the European Commission.

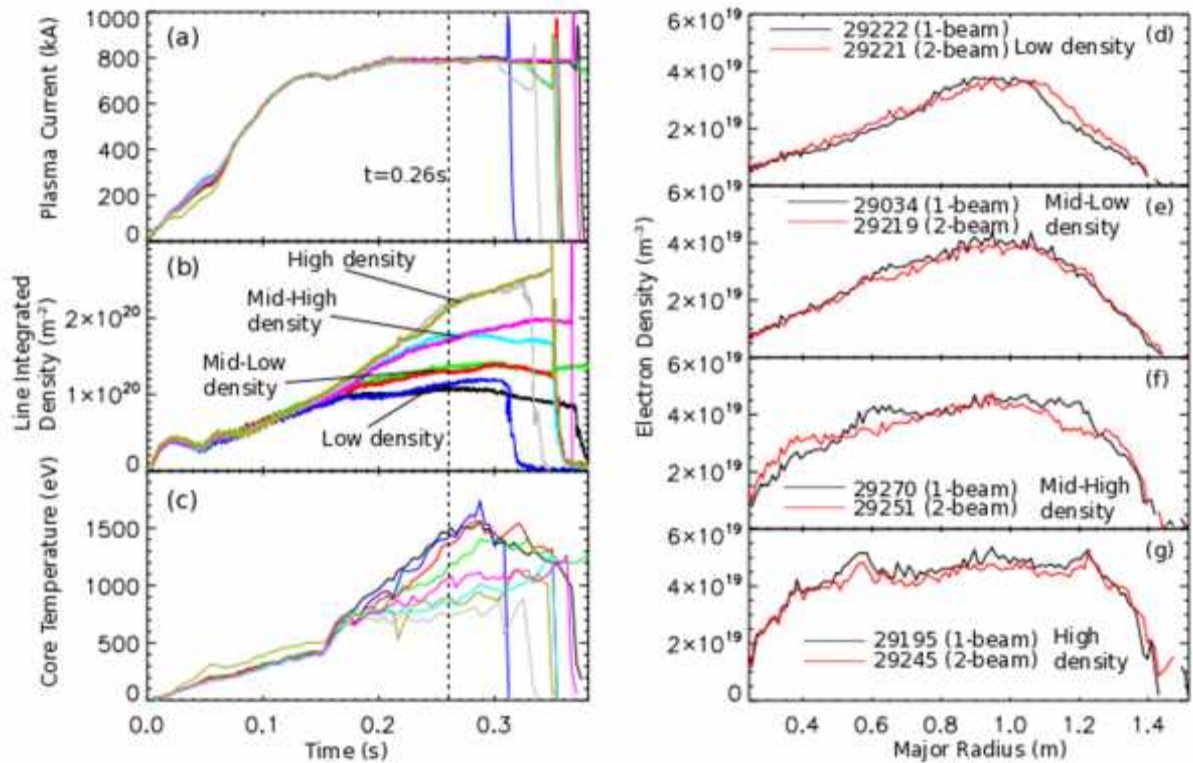

Figure 1: Time traces of a) plasma current, b) interferometer measurements of line-integrated density and c) Thomson scattering measurements of core temperature for the 8 shots comprising the power/density scan. d)-g): Thomson scattering density profiles taken at t=0.26s of the 1-beam (black) and 2-beam (red) shots at each density level.

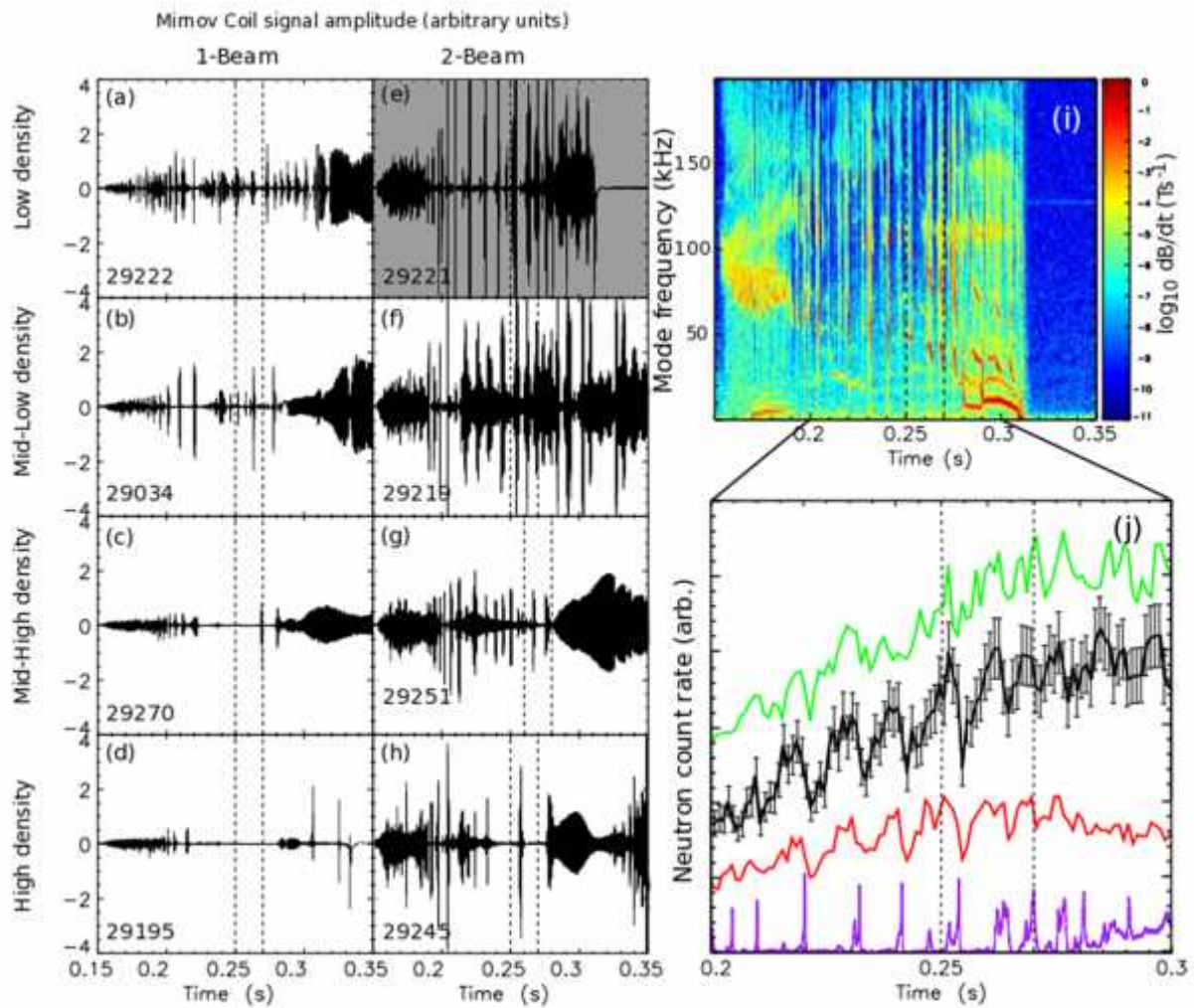

Figure 2: a)-h) Time traces from a Mirnov coil situated on the outboard side of MAST for the density/power scan with the Period-of-Interest for each shot marked by vertical dashed lines.

(i) Magnetics spectrogram of shot 29221 (low density, 2-beam indicated in grey in fig 2(e)) illustrating the presence of n=1 fishbone modes during the Period-Of-Interest (0.25-0.27s, indicated).

(j) Neutron emission rates for shot 29221. Green: Fission Chamber measurement, Black: NC midplane inboard chord (impact parameter=0.948±0.03m), Red: NC midplane outboard chord (impact parameter=1.14±0.03m), Purple: R.M.S. of Mirnov coil signal. Drops in neutron emission rate are very well correlated with FB modes, particularly in the NC signals. Error bars indicating a 10% relative error are shown for one of the NC chords [22] which are of a similar magnitude to errors in the F.C. signal (8.5% relative magnitude, omitted for clarity).

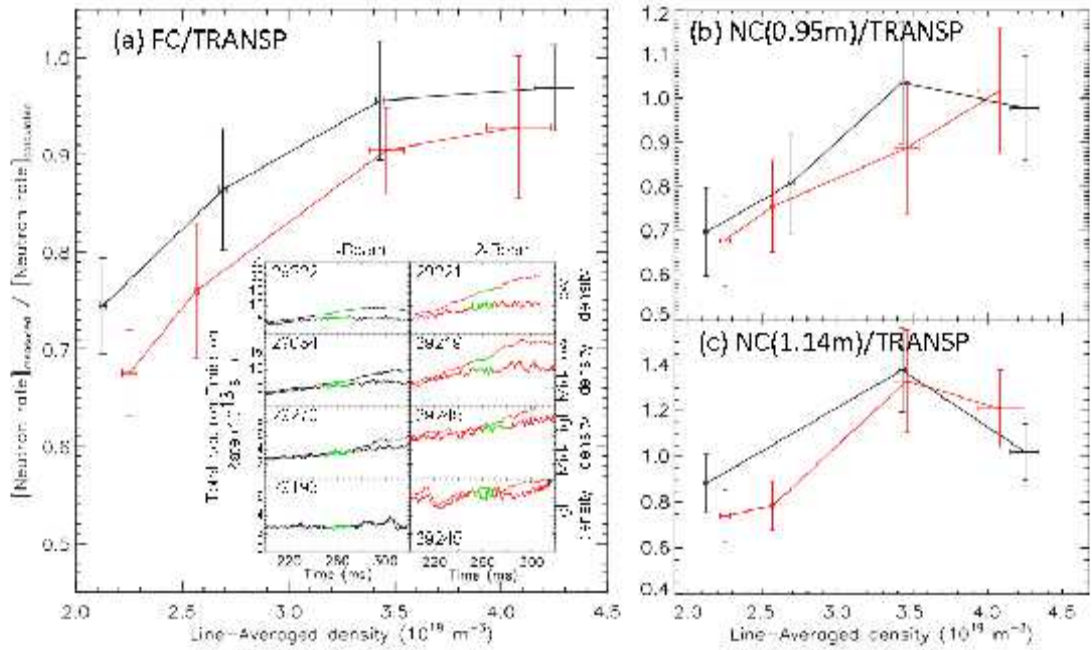

Figure 3: (a) (inset) neutron emission rate measured by the FC (solid) and calculated by TRANSP (dashed) with indicated POI (green) for each time trace. (a) (Main) FC measurement divided by TRANSP calculated neutron emission rate averaged over the POI plotted versus line-averaged density. Black =1-beam shots, red=2-beam shots. (b) NC measured neutron count rate divided by TRANSP calculated neutron rate averaged over the POI for NC line of sight with impact parameter =0.95m. (c) NC measured neutron count rate divided by TRANSP calculated neutron rate averaged over the POI for NC line of sight with impact parameter =1.14m.

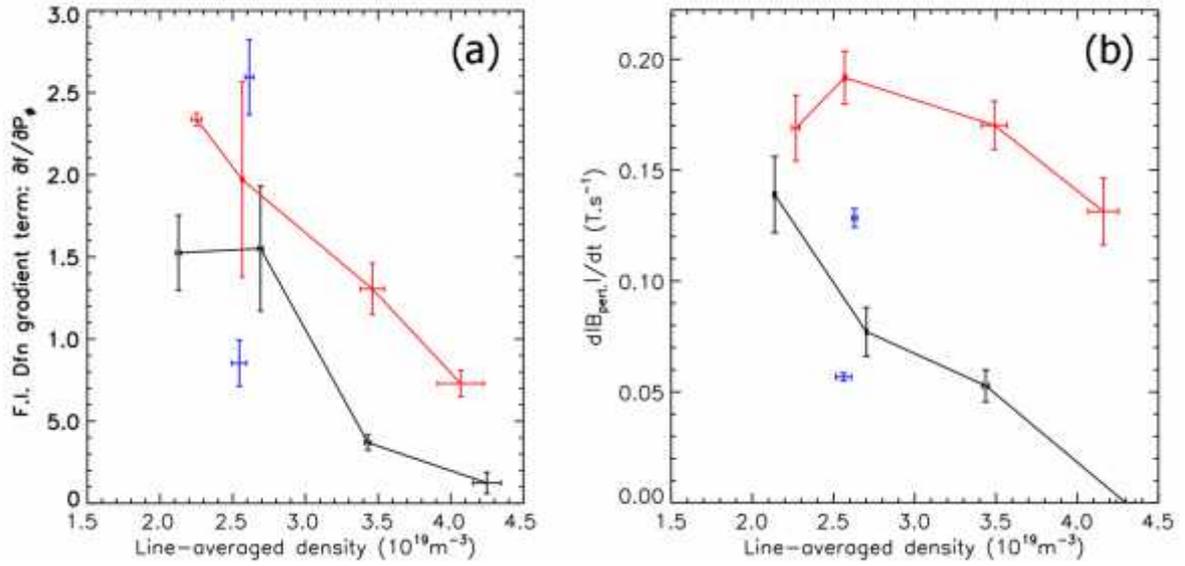

Figure 4: (a) F.I. distribution function gradient $\partial f/\partial P_\varphi$ (plotted with arbitrary units) averaged over POI. Black: 1-beam shots, red: 2-beam shots. Also shown for comparison are the 1 and 2-beam shots carried out in previous experiments reported in [6]. (b) Mode growth rate $d|B_{pert.}|/dt$ versus line-averaged density for Black: 1-beam shots, Red: 2-beam shots, Blue: 1 and 2-beam shots carried out in previous experiments reported in [6].

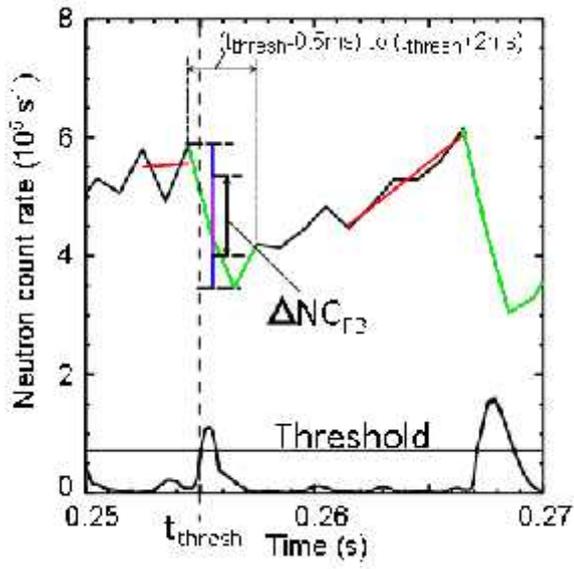

Figure 5: Illustration of the analysis method to determine the quantity $\Delta NC_{FB}$. See text for details.

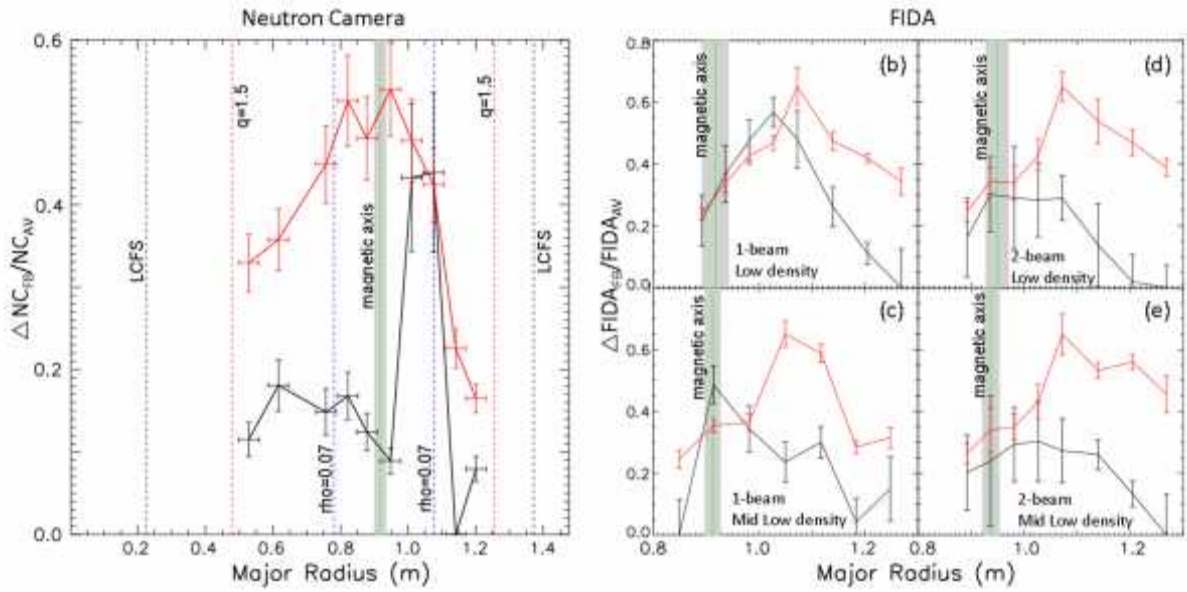

Figure 6: (a) relative drop in neutron rate due to fishbones as a function of NC impact parameter (solid black line and data points), normalised neutron emission rate (solid red line and data points, arbitrary scale). Representative quantities from MSE constrained EFIT are shown as follows (in the case of flux surface quantities the midplane major radius is shown): Magnetic axis (vertical green dashed line), $\psi_{P, normalised}$ =0.07 (vertical blue dotted line), q=1.5 (vertical red dash-dotted line), LCFS (vertical black dashed line). MAST pulses used in obtaining profile: 29222, 29924, 29928, 29929, 29931.

(b-e) Relative drop in FIDA emission due to FB modes for the 1-beam and 2-beam shots at the 2 lower density levels in the scan (black line and data points). FIDA emission (red line and data points, arbitrary scale). Also shown is the position of the magnetic axis from MSE constrained EFIT.

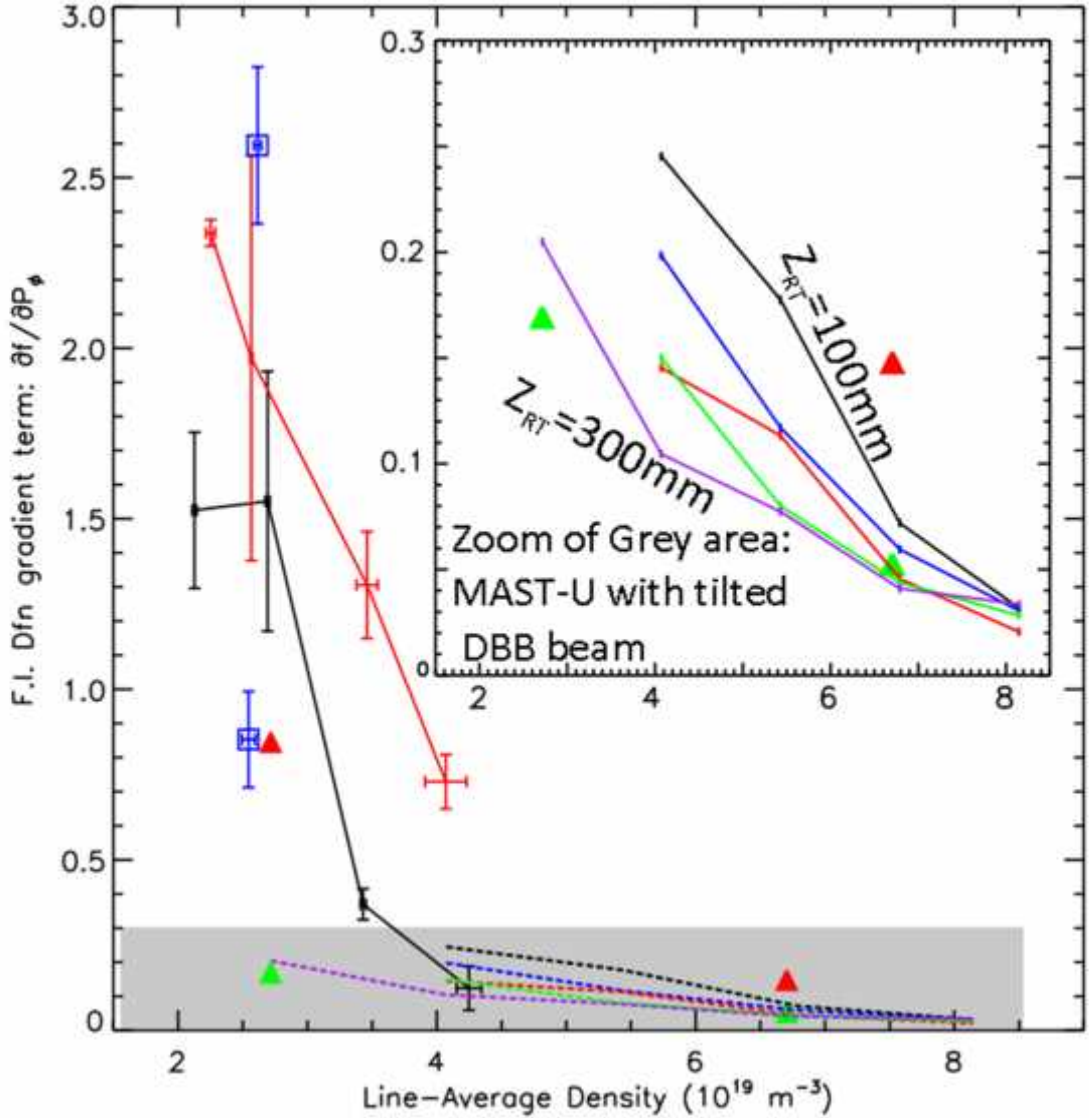

Figure 7: maximum gradient $\partial f/\partial P_\varphi$ (plotted with arbitrary units, as in Fig. 4a) versus line averaged density for 1-beam shots(black), 2-beam shots(red). These lines are identical to those shown in figure 4a. Also shown are the MAST-Upgrade core scope (green triangles) and higher power 3-beam simulations representing future upgrades beyond Core Scope (red triangles). 1-beam and 2-beam shots from previous experiments (blue squares, see [6] for details). Dashed lines: MAST-Upgrade stage 1 with various injection geometries for the nominally on-axis DBB beam with 100mm<$Z_{RT}$<300mm (See text for further details).

Inset: Zoom plot of grey region showing more than a factor of 2 difference in $\partial f/\partial P_\varphi$ at density ~4×10$^{19}$m$^{-3}$ for the extremes of the range of $Z_{RT}$ investigated.